\documentclass[12pt, draftclsnofoot, onecolumn]{IEEEtran}
\usepackage{color}
\usepackage{leftidx}
\usepackage{cite}
\usepackage{microtype}
\usepackage{cleveref}
\usepackage{graphicx,amssymb,amstext,amsmath}
\usepackage[ruled,vlined,lined,ruled,linesnumbered]{algorithm2e}
\usepackage{subfigure}
 \usepackage{floatrow}
\usepackage[font=small,skip=5pt]{caption}

\begin{document}

\title{{\fontsize{22pt}{1em}\selectfont Computation Rate Maximization for Wireless Powered Mobile-Edge Computing with Binary Computation Offloading}}

\author{Suzhi~Bi and Ying-Jun Angela Zhang\\
\thanks{S.~Bi (bsz@szu.edu.cn) is with the College of Information Engineering, Shenzhen University, Shenzhen, China. Y-J.~A.~Zhang (yjzhang@ie.cuhk.edu.hk) is with the Department of Information Engineering, The Chinese University of Hong Kong, HK.}
\vspace{-8ex}}

\maketitle

\IEEEpeerreviewmaketitle

\begin{abstract}
Finite battery lifetime and low computing capability of size-constrained wireless devices (WDs) have been longstanding performance limitations of many low-power wireless networks, e.g., wireless sensor networks (WSNs) and Internet of Things (IoT). The recent development of radio frequency (RF) based wireless power transfer (WPT) and mobile edge computing (MEC) technologies provide promising solutions to fully remove these limitations so as to achieve sustainable device operation and enhanced computational capability. In this paper, we consider a multi-user MEC network powered by WPT, where each energy-harvesting WD follows a binary computation offloading policy, i.e., data set of a task has to be executed as a whole either locally or remotely at the MEC server via task offloading. In particular, we are interested in maximizing the (weighted) sum computation rate of all the WDs in the network by jointly optimizing the individual computing mode selection (i.e., local computing or offloading) and the system transmission time allocation (on WPT and task offloading). The major difficulty lies in the combinatorial nature of multi-user computing mode selection and its strong coupling with transmission time allocation. To tackle this problem, we first consider a decoupled optimization, where we assume that the mode selection is given and propose a simple bi-section search algorithm to obtain the conditional optimal time allocation. On top of that, a coordinate descent method is devised to optimize the mode selection. The method is simple in implementation but may suffer from high computational complexity in a large-size network. To address this problem, we further propose a joint optimization method based on the ADMM (alternating direction method of multipliers) decomposition technique, which enjoys much slower increase of computational complexity as the networks size increases. Extensive simulations show that both the proposed methods can efficiently achieve near-optimal performance under various network setups, and significantly outperform the other representative benchmark methods considered.
\end{abstract}
\vspace{-3ex}


\vspace{-3ex}

\section{Introduction}
The recent development of Internet of Things (IoT) technology is a key step towards truly intelligent and autonomous control in many important industrial and commercial systems, such as smart power grid and smart home automation \cite{2015:Fuqaha}. In an IoT network, massive number of wireless devices (WDs) capable of communication and computation are deployed. Due to the stringent device size constraint and production cost consideration, an IoT device (e.g., sensor) often carries a capacity-limited battery and an energy-saving low-performance processor. As a result, the \emph{finite device lifetime} and \emph{low computing capability} are unable to support increasingly many new applications that require sustainable and high-performance computations, e.g., autonomous driving and augmented reality. Therefore, how to tackle the two fundamental performance limitations is a critical problem in the research and development of modern IoT technology.

Recently, radio frequency (RF) based \emph{wireless power transfer} (WPT) has emerged as an effective solution to the finite battery capacity problem \cite{2015:Bi,2015:Lu,2016:Bi}. Specifically, WPT uses dedicated RF energy transmitter, which can continuously charge the battery of remote energy-harvesting devices. Currently, commercial WPT transmitter can effectively deliver tens of microwatts RF power to a distance of more than $10$ meters, which is sufficient to power the activities of many low-power WDs \cite{2016:Bi1}. Meanwhile, we expect much more efficient WPT in the near future, considering the fast development of WPT circuit design and advanced signal processing techniques, e.g., energy beamforming \cite{2013:Zhang,2015:Zeng} and distributed multi-point WPT \cite{2016:Bi2}. The application of WPT to power wireless communication devices has attracted extensive research interests \cite{2016:Bi1,2014:Ju,2014:Liu}. Thanks to the broadcasting nature of RF signal, WPT is particularly suitable for powering a large number of closely-located WDs, like those deployed in WSNs and IoT.

On the other hand, a recent technology innovation named \emph{mobile edge computing} (MEC) is proposed as a cost-effective method to enhance the computing capability of wireless devices \cite{2016:Chiang,2017:Mao}. As its name suggests, MEC allows the WDs to offload intensive computations to nearby servers located at the edge of radio access network, e.g., cellular base station and WiFi access point (AP). Compared with the conventional cloud computing paradigm, MEC removes long backhaul latency, and enjoys lower device energy consumption and superior server load balancing performance. In particular, MEC hits a perfect match with the IoT technology, and thus has attracted massive investment from many major technology companies, such as Huawei, Intel and IBM, and has been identified as a key technology towards future 5G network \cite{2015:ETSI}. In general, there are two basic computation task offloading models in MEC, i.e., binary and partial computation offloading \cite{2017:Mao}. Specifically, \emph{binary offloading} requires a task to be executed as a whole either locally at the WD or remotely at the MEC server. Partial offloading, on the other hand, allows a task to be partitioned into two parts with one executed locally and the other offloaded for edge execution. In practice, binary offloading is easier to implement and suitable for simple tasks that are not partitionable, while partial offloading is favorable for some complex tasks composed of multiple parallel segments.

\begin{figure}
\centering
  \begin{center}
    \includegraphics[width=0.6\textwidth]{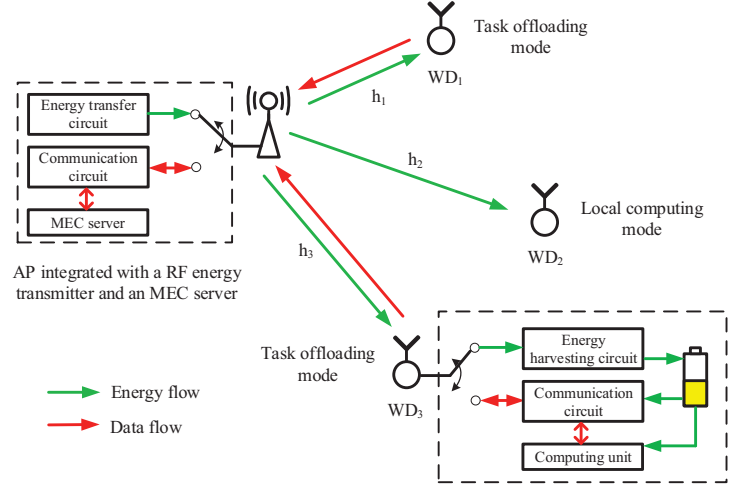}
  \end{center}
  \caption{An example $3$-user wireless powered MEC system with binary computation offloading.}
  \label{101}
\end{figure}

In conventional battery-powered MEC networks, a key research problem is the joint design of task offloading and system resource allocation to optimize the computing performance \cite{2013:Wu,2016:Wang,2017:You,2016:Chen}. For a single-user MEC, \cite{2013:Wu} studies the optimal binary offloading decision to minimize the energy consumption under stochastic wireless channel, where it optimizes the CPU frequency in local computing mode and the transmission data rate in offloading mode. For partial offloading mode, \cite{2016:Wang} jointly optimizes the offloading ratio, transmission power and CPU frequency to either minimize energy consumption or computation latency. For multi-user MEC with partial offloading, \cite{2017:You} allows the users to share the MEC server in time and formulates a convex optimization to minimize the weighted sum energy consumption of the users by jointly optimizing the offloading ratio and time. Multi-user MEC with binary offloading is a more complicated scenario, which often involves non-convex combinatorial optimization problems. In \cite{2016:Chen}, a heuristic algorithm based on separable semidefinite relaxation is proposed to optimize binary offloading decisions and wireless resource allocation for minimum energy consumptions.

The integration of WPT and MEC technologies introduces a new paradigm named \emph{wireless powered MEC}, which can potentially tackle the two fundamental performance limitations in IoT networks. Meanwhile, it brings new challenges to the optimal system design. On one hand, the task offloading and resource allocation decisions in MEC now depend on the distinct amount of energy harvested by individual WDs from WPT. On the other hand, WPT and task offloading need to share the limited wireless resource, e.g., time or frequency. There are few existing studies on wireless powered MEC system \cite{2016:You,2017:Wang1,2017:Wang2}. \cite{2016:You} considers a single-user wireless powered MEC with binary offloading, where the user maximizes its probability of successful computation under latency constraint. In a multi-user scenario, \cite{2017:Wang1} considers using a multi-antenna AP to power the users and minimizes the AP's total energy consumption subject to the users' individual latency constraints. A closely related work to this paper is \cite{2017:Wang2}, which maximizes the weighted sum computation rate of a multi-user wireless powered MEC network. However, both \cite{2017:Wang1} and \cite{2017:Wang2} assume partial computation offloading policy. In contrast, the optimal design of binary offloading policy, which is widely adopted in IoT networks by simple computing tasks, is currently lacking of study. Mathematically speaking, partial offloading is a convex-relaxed version of the binary offloading policy, which avoids the hard combinatorial mode selection problem in system design. In fact, both \cite{2017:Wang1} and \cite{2017:Wang2} derived convex optimization formulations, such that the optimal solution can be efficiently obtained with off-the-shelf algorithms. The optimal design under the binary offloading policy in a multi-user environment, however, is a much more challenging problem, which even has not been fully addressed in conventional battery-powered MEC.

In this paper, we consider a wireless powered MEC network as shown in Fig.~\ref{101}, where the AP is reused as both energy transmitter and MEC server that transfers RF power to and receives computation offload from the WDs. Each device follows the \emph{binary offloading policy}. In particular, we are interested in maximizing the \emph{weighted sum computation rate}, i.e., the number of processed bits per second, of all the WDs in the network, which is a direct measure of the overall computing capability of the system \cite{2017:Wang2}. To the authors' best knowledge, this is the first paper that studies the optimal design in a multi-user wireless powered MEC network using binary computation offloading policy. Our contributions are detailed below.
\begin{enumerate}
  \item We formulate the problem as a joint optimization of individual computing mode selection (i.e., offloading or local computing) and the system transmission time allocation (on WPT and task offloading). The combinatorial nature of multi-user computing mode selection and its strong coupling with time allocation make the optimal solution hard to obtain in general. As a performance benchmark, a mode enumeration-based optimal method is presented for evaluating the other reduced-complexity algorithms proposed in this paper.
  \item We first propose a decoupled optimization method. With a given mode selection decision, we derive a semi-closed-form solution of the optimal time allocation. Then, we propose a simple bi-section search algorithm that can efficiently obtain the optimal time allocation. On top of that, a coordinate descent (CD) method is devised to optimize the mode selection. The method is simple in implementation as it involves only basic function evaluations. However, the overall computational complexity grows like $O(N^3)$, where $N$ is the network size. As such, the computational complexity may become undesirable when $N$ is too large.
  \item To address the complexity issue in large-size networks, we further devise an ADMM-based technique that jointly optimizes the mode selection and time allocation. The proposed method tackles the hard combinatorial mode selection by decomposing the original problem into parallel small-scale integer programming subproblems, one for each WD. Compared to the CD method, the ADMM-based method requires more complex calculations, e.g., projected Newton's method \cite{2004:Boyd}. On the other hand, its computational complexity increases much more slowly at a linear rate $O(N)$ of the network size.
\end{enumerate}

Extensive simulations show that both the proposed algorithms can achieve \emph{near-optimal} performance under various network setups, and significantly outperform the other benchmark algorithms, e.g., the convex relaxation method. In practice, based on their respective features, it is more preferable to apply the CD method when network size is small (e.g., $\leq 30$ WDs) or the AP is hardware-constrained, and to use ADMM-based method in a large-size network where the impact of network size dominates the overall computational complexity. Interestingly, in a special case where all the WDs are of equal computation energy efficiency and weight, we observe that the optimal computing mode selection has a \emph{threshold structure} based on the wireless channel strength. Accordingly, the optimal computing mode can be easily obtained by searching the threshold from the WD with the strongest channel to the weakest.

The rest of the paper is organized as follows. In Section II, we introduce the system model of the wireless powered MEC. The computation rate maximization problem is formulated in Section III. In Section IV and V, we propose two efficient algorithms to solve the problem with different practical features. In Section VI, we discuss some practical extensions of the proposed algorithms. In Section VII, simulation results are presented to evaluate the proposed algorithms. Finally, we conclude the paper and discuss future directions in Section VIII.

\section{System Model}
\subsection{Network Model}
As shown in Fig.~\ref{101}, we consider a wireless powered MEC network consisting of an AP and $N$ WDs, where the AP and the WDs have a single antenna each. In particular, an RF energy transmitter and a MEC server is integrated at the AP. The AP is assumed to be connected to a stable power supply and broadcast RF energy to the distributed WDs, while each WD has an energy harvesting circuit and a rechargeable battery that can store the harvested energy to power its operations. Each device, including the AP and the WDs, has a communication circuit. Specifically, we assume that WPT and communication are performed in the same frequency band. To avoid mutual interference, the communication and energy harvesting circuits of each WD operate in a time-division-multiplexing (TDD) manner. A similar TDD circuit structure is also applied at the AP to separate energy transmission and communication with the WDs. Within each system time frame of duration $T$, the wireless channel gain between the AP and the $i$-th WD is denoted by $h_i$, which is assumed reciprocal for the downlink and uplink,\footnote{The channel reciprocity assumption is made to obtain more design insights on the impact of wireless channel conditions. The proposed algorithms in this paper, however, can be easily extended to the case with non-equal uplink and downlink channels.} and static within each time frame but may vary across different time frames.

Within each time frame, we assume that each WD needs to accomplish a certain computing task based on its local data. For instance, a WD as a wireless sensor needs to regularly generate an estimate, e.g., the pollution level of the monitored area, based on the raw data samples measured from the environment. In particular, the computing task of a WD can be performed locally by the on-chip micro-processor, which has low computing capability due to the energy- and size-constrained computing processor. Alternatively, the WD can also offload the data to the MEC server with much more powerful processing power, which will compute the task and send the result back to the WD.

In this paper, we assume that the WDs adopt a binary computation offloading rule. That is, a WD must choose to operate in either the local computing mode (mode $0$, like WD$_2$ in Fig.~1) or the offloading mode (mode $1$, like WD$_1$ and WD$_3$) in each time frame. In practice, this corresponds to a wide variety of applications. For instance, the measurement samples of a sensor are correlated in time, and thus need to be jointly processed to enhance the estimation accuracy.

\begin{figure}
\centering
  \begin{center}
    \includegraphics[width=0.6\textwidth]{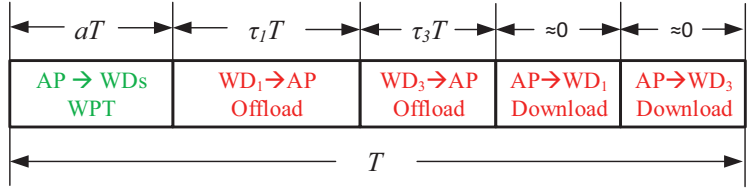}
  \end{center}
  \caption{An example time allocation in the $3$-user wireless powered MEC network in Fig.~\ref{101}. Only WD$_1$ and WD$_3$ selecting mode $1$ offload the task to and download the computation results from the AP.}
  \label{110}
\end{figure}

\subsection{Computation Model}
We consider an example transmission time allocation in Fig.~\ref{110}. We use two mutually exclusive sets $\mathcal{M}_0$ and $\mathcal{M}_1$ to denote the indices of WDs that operate in mode $0$ and $1$, respectively. As such $\mathcal{M} = \mathcal{M}_0 \cup \mathcal{M}_1 = \{1,\cdots,N\}$ is the set of all the WDs. In the first part of a tagged time frame, the AP broadcasts wireless energy to the WDs for $aT$ amount of time, where $a\in[0,1]$, and all the WDs harvest the energy. Specifically, the energy harvested by the $i$-th WD is
\begin{equation}
\label{103}
E_i = \mu P h_i a T,\ i=1,\cdots,N,
\end{equation}
where $P$ denotes the RF energy transmit power of the AP and $\mu\in(0,1)$ denotes the energy harvesting efficiency \cite{2013:Zhang}. In the second part of the time frame $(1-a)T$, the WDs in $\mathcal{M}_1$ (e.g., WD$_1$ and WD$_3$ in Fig.~\ref{101}) offload the data to the AP. To avoid co-channel interference, we assume that the WDs take turns to transmit in the uplink, and the time that a WD$_i$ transmits is denoted by $\tau_i T$, $\tau_i\in [0,1]$. Depending on the selected computing mode, the detailed operation of each WD is illustrated as follows.

\subsubsection{Local Computing Mode}
Notice that the energy harvesting circuit and the computing unit are separate. Thus, a mode-$0$ WD can \emph{harvest energy and compute its task simultaneously} \cite{2017:Wang1}. That is, it can compute throughout the entire time frame of duration $T$. Let $\phi>0$ denote the number of computation cycles needed to process one bit of raw data, which is determined by the nature of the application and is assumed to be equal for all the WDs. Let $f_i$ denote the processor's chosen computing speed (cycles per second) and $0\leq t_i\leq T$ denote the computation time of the WD. $f_i\leq f_{max}$ holds as the computation capability constraint. The power consumption of the processor is modeled as $k_i f_i^3$ (joule per second), where $k_i$ denotes the computation energy efficiency coefficient of the processor's chip \cite{2016:Wang}. Then, the total energy consumption is constrained by
\begin{equation}
\label{2}
k_i f_i^3 t_i \leq E_i
\end{equation}
to ensure sustainable operation of the WD.\footnote{We assume each WD has sufficient initial energy in the very beginning and the battery capacity is sufficiently large such that battery-overcharging is negligible.} In particular, we assume that the WDs are \emph{energy-constrained}, such that a WD can always consume all the harvested energy within a time frame by operating at the maximum computing speed. In other words,
\begin{equation}
\label{132}
E_i = \mu P h_i a T\leq \mu P h_i T < k_i f_{max}^3 T
\end{equation}
holds for any practical value of $h_i$ and $i=1,\cdots,N$. Accordingly, the computation rate of a mode-$0$ WD$_i$ (in bits per second), denoted by $r_{L,i}$, can be calculated as \cite{2016:Wang}
\begin{equation}
\label{1}
r_{L,i} = \frac{f_i t_i}{\phi T},\ \forall i \in \mathcal{M}_0.
\end{equation}

\subsubsection{Offloading Mode}
Due to the TDD circuit constraint, a mode-$1$ WD can only \emph{offload its task to the AP after harvesting energy}. We denote the number of bits to be offloaded to the AP as $v_u b_i$, where $b_i$ denotes the amount of raw data and $v_u>1$ indicates the communication overhead in task offloading, such as packet header and encryption. Let $P_i$ and $\tau_i T$ denote the transmit power and time of the $i$-th WD, respectively. Then, the maximum $b_i^*$ equals to the data transmission capacity, i.e.,
\begin{equation}
\label{3}
b_i^* =  \frac{B\tau_i T}{v_u}\log_2\left(1+\frac{P_i h_i}{ N_0}\right), \ \forall i\in \mathcal{M}_1,
\end{equation}
where $B$ denotes the communication bandwidth and $N_0$ denotes the receiver noise power.

After receiving the raw data of all the WDs, the AP computes and sends back the output result of length $r_d b_i$ bits back to the corresponding WD. Here, $r_d \ll 1$ indicates the output/input ratio including the overhead in downlink transmission. Let $f_0$ denote the AP processor's fixed computing speed and $P_0$ denote the transmit power of the AP. The time spent on task computation and feeding back to WD$_i$ is
\begin{equation}
\label{141}
l_i = \frac{\phi b_i}{f_0} + \frac{r_d b_i}{B \log_2\left(1+\frac{P_0 h_i}{N_0}\right)}.
\end{equation}
In practice, the computing capability and the transmit power of the AP is much stronger than the energy-harvesting WDs, e.g., by more than three orders of magnitude. Beside, $r_d$ is a very small value, e.g., one output temperature estimation from tens of input sensing sample. Accordingly, we can infer from (\ref{3}) and (\ref{141}) that $l_i \ll \tau_i T$, and thus the time spent on task computation and result feedback by the AP can be safely neglected like in \cite{2013:Wu,2016:You,2017:Wang1}. In this case, task offloading can occupy the rest of the time frame after WPT, i.e.,
\begin{equation}
\label{11}
\sum_{i\in \mathcal{M}_1} \tau_i + a \leq 1.
\end{equation}
Besides, from the above discussion, we also neglect the energy consumption by the WD on receiving the computation result from the AP, and consider only the energy consumptions on data transmission to the AP.\footnote{The energy and time consumed on channel estimation and coordination can be modeled as two constant terms that will not affect the validity of the proposed algorithms. For simplicity of illustration, they are neglected in this paper.} In this case, a WD needs to exhaust its harvested energy on task offloading to maximize the computation rate, i.e., $P_i^* = \frac{E_i}{\tau_i T}$.\footnote{Same as most of the existing work on wireless powered communications, e.g., \cite{2014:Ju,2014:Liu}, we do not assume a maximum transmit power constraint for the WDs because of the small amount energy harvested from WPT in practice.}  By substituting $P_i^*$ into (\ref{3}), the maximum computation rate of a mode-$1$ WD$_i$, denoted by $r_{O,i}^*$, can be expressed as
\begin{equation}
\label{4}
r_{O,i}^* = \frac{b_i^*}{T} = \frac{B\tau_i}{v_u}\log_2\left(1+\frac{\mu P a h_i^2}{ \tau_i N_0}\right), \ \forall i\in \mathcal{M}_1.
\end{equation}
In the next section, we formulate the weighted sum-rate maximization problem of the considered wireless powered MEC system.

\section{Problem Formulation}
In this paper, we maximize the weighted sum computation rate of all the WDs in each time frame. From (\ref{1}) and (\ref{4}), we can see that the computation rates of the WDs are related to their computing mode selection and the system resource allocation on WPT, communication, and computation. Mathematically, the computation rate maximization problem is formulated as follows.
 \begin{subequations}
   \begin{align}
    (P1):\ \ \ & \underset{\mathcal{M}_0, a, \boldsymbol{\tau},\mathbf{f,t}}{\text{maximize}} & &  \sum_{i\in \mathcal{M}_0} w_i\frac{f_i t_i}{\phi T}  +  \sum_{j \in \mathcal{M}_1} w_j \frac{B\tau_j}{v_u}\log_2\left(1+\frac{\mu P a h_j^2}{ \tau_j N_0}\right) \label{61}\\
    & \text{subject to} &  & \sum_{j\in\mathcal{M}_1} \tau_j + a \leq 1, \label{62}\\
    & & & k_i f_i^3 t_i \leq \mu P h_i a T, \ \forall i\in \mathcal{M}_0 , \label{63}\\
    & & & 0\leq t_i \leq T,\ 0 \leq f_i \leq f_{max},\ \forall i\in \mathcal{M}_0, \label{64}\\
    &  & & a \geq 0,\ \tau_j \geq 0, \ \forall j \in \mathcal{M}_1, \\
    & & & \mathcal{M}_0  \subseteq \mathcal{M}, \ \mathcal{M}_1 = \mathcal{M}\setminus  \mathcal{M}_0.
   \end{align}
   \label{6}
\end{subequations}
Here, $w_i>0$ denotes the weight of the $i$-th WD. $\mathbf{f}=\left\{f_i| i\in\mathcal{M}_0\right\}$ and $\mathbf{t}=\left\{t_i| i\in\mathcal{M}_0\right\}$ denote the computing speed and computation time of the mode-$0$ WDs. $\boldsymbol{\tau}= \left\{\tau_j| j\in\mathcal{M}_1\right\}$ denotes the offloading time of the mode-$1$ WDs. The two terms of the objective function correspond to the computation rates of mode-$0$ and mode-$1$ WDs, respectively. (\ref{62}) is the time allocation constraint and (\ref{63}) denotes the individual energy harvesting constraints for mode-$0$ WDs.

Problem (P1) is evidently non-convex due to the combinatorial mode selection variable $\mathcal{M}_0$ and the multiplicative terms in both the objective function and constraints. A close observation of (P1) shows that we can independently optimize the computing speed $f_i$ and duration $t_i$ of each mode-$0$ WD$_i$ without affecting the performance of the other WDs, when the WPT time $aT$ is fixed. Specifically, we have the following lemma on the maximum local computation rate.

\textbf{Lemma $1$:} The maximum $r_{L,i}^* = $ is achieved when $t_i^* =T$ and $f_i^*=\left(\frac{E_i}{k_i T}\right)^{\frac{1}{3}}$.

\emph{Proof}: \ For a tagged mode-$0$ WD$_i$, we have $r_{L,i} = \frac{f_i t_i}{\phi T} \leq \frac{1}{\phi T}\left(\frac{E_i}{k_i}\right)^{\frac{1}{3}}t_i^{\frac{2}{3}}$, where the inequality is obtained from (\ref{2}) and the upper bound is achievable by exhausting all the harvested energy on computation. Note that $r_{L,i}$ increases monotonically with $t_i$. Hence, the maximum $r_{L,i}^*$ is achieved by setting $t_i^*=T$, i.e., the WD computes for a maximal allowable time throughout the time frame. Accordingly, we have from (\ref{2}) that $f_i^* = \min\left(\left(\frac{E_i}{k_i T}\right)^{\frac{1}{3}},f_{max}\right)$. By the assumption in (\ref{132}), we can infer that $\left(\frac{E_i}{k_i T}\right)^{\frac{1}{3}}< f_{max}$ always holds. Thus, $f_i^* = \left(\frac{E_i}{k_i T}\right)^{\frac{1}{3}}$.  $\hfill \blacksquare$

By substituting $t_i^*=T$ and $f_i^*=\left(\frac{E_i}{k_i T}\right)^{\frac{1}{3}}$ into (\ref{1}), the maximum local computation rate
\begin{equation}
\label{114}
r_{L,i}^*  = \frac{f_i^* t_i^*}{\phi T} =\eta_1 \left(\frac{h_i}{k_i}\right)^{\frac{1}{3}} a^{\frac{1}{3}},\ \forall i\in \mathcal{M}_0,
\end{equation}
where $\eta_1 \triangleq  \frac{\left( \mu P \right)^{\frac{1}{3}}}{\phi}$ is a fixed parameter. Accordingly, we can replace the first term in (P1) with the RHS of (\ref{114}) to safely remove the variables $\mathbf{f}$, $\mathbf{t}$ and the corresponding constraints in (\ref{63}) and (\ref{64}). This yields an equivalent simplification of (P1):
 \begin{subequations}
   \begin{align}
    (P2): \ \ & \underset{\mathcal{M}_0,a, \boldsymbol{\tau}}{\text{maximize}} & &  \sum_{i \in \mathcal{M}_0} w_i  \eta_1 \left(\frac{h_i}{k_i}\right)^{\frac{1}{3}} a^{\frac{1}{3}} +  \sum_{j \in \mathcal{M}_1}  w_j \varepsilon \tau_j\ln \left(1+ \frac{\eta_2 h_j^2 a}{\tau_j}\right) \\
    & \text{subject to} &  & \sum_{j \in \mathcal{M}_1} \tau_j + a \leq 1, \label{116}\\
    & & & a \geq 0,\ \tau_j\geq 0, \ \forall j\in \mathcal{M}_1,\ \mathcal{M}_0  \subseteq \mathcal{M}, \ \mathcal{M}_1 = \mathcal{M}\setminus  \mathcal{M}_0,
   \end{align}
\end{subequations}
where $\eta_2 \triangleq \frac{\mu P}{N_0}$ and $\varepsilon \triangleq \frac{B}{v_u \ln 2}$. Among all the parameters in (P2), only the wireless channel gains $h_i$'s are time-varying in each time frame within the considered period, while the others (e.g., $w_i$'s and $k_i$'s) are assumed to remain constant.

(P2) is still a hard non-convex problem due to the combinatorial computing mode selection. However, we observe that the second term in the objective is jointly concave in $(a,\tau_j)$. Once $\mathcal{M}_0$ is given, (P2) reduces to a convex problem, where the optimal time allocation $\left\{a^*,\boldsymbol{\tau}^*\right\}$ can be efficiently solved using off-the-shelf optimization algorithms, e.g., interior point method \cite{2004:Boyd}. Accordingly, a straightforward method is to enumerate all the $2^N$ possible $\mathcal{M}_0$ and output the one that yields the highest objective value. The enumeration-based method may be applicable for a small number of WDs, e.g., $N\leq 10$, however, quickly becomes computationally infeasible as $N$ further increases. Therefore, it will be mainly used as a benchmark to evaluate the performance of the proposed reduced-complexity algorithms in this paper.

\section{Decoupled Optimization using Coordinate Descent Method}\label{decouple}
In this section, we propose a decoupled optimization method, where we first assume that $\mathcal{M}_0$ is given and derive a semi-closed-form expression of the optimal time allocation $\left\{a^*,\boldsymbol{\tau}^*\right\}$. Subsequently, a low-complexity bi-section search can be applied to obtain the optimal solution. On top of that, we then devise a coordinate descent method that optimizes the mode selection. In addition, we further study a homogeneous special case, where the WDs have equal weight and computing efficiency, and obtain some interesting insights of the optimal solution.

\subsection{Optimal Transmission Time Allocation Given $\mathcal{M}_0$}
In this subsection, we study the optimal transmission time allocation problem in (P2) given $\mathcal{M}_0$. In particular, we propose a simple bi-section search algorithm that has much lower complexity than general convex optimization techniques, e.g., interior point method. Besides, interesting design insights are obtained from the analysis in this subsection.

Suppose that $\mathcal{M}_0$ is given in (P2). Let us introduce a Lagrangian multiplier to constraint (\ref{116}) to form a partial Lagrangian
 \begin{equation}
 \label{14}
   \begin{aligned}
    & L(a,\boldsymbol{\tau},\nu) = \sum_{i \in \mathcal{M}_0} w_i  \eta_1 \left(\frac{h_i}{k_i}\right)^{\frac{1}{3}} a^{\frac{1}{3}} +  \sum_{j \in \mathcal{M}_1} w_j \varepsilon \tau_j\ln \left(1+ \frac{\eta_2 h_j^2 a}{\tau_j}\right)+ \nu \left(1-a -\sum_{j\in \mathcal{M}_1} \tau_j\right).
   \end{aligned}
\end{equation}
The corresponding dual function is
\begin{equation}
d(\nu) = \underset{a,\boldsymbol{\tau}}{\text{maximize}}\ \left\{L\left(a,\boldsymbol{\tau},\nu\right)\mid a \geq 0,\ \tau_j\geq 0, \ \forall j\in \mathcal{M}_1\right\},
\end{equation}
and the dual problem is
\begin{equation}
\underset{\nu }{\text{minimize}}\ \left\{d\left(\mathbf{\nu}\right)\mid \nu\geq 0 \right\}.
\end{equation}
As (P2) is a convex problem given $\mathcal{M}_0$, the dual problem achieves the same optimal objective value by the strong duality. It can be seen that equation $\sum_{j\in \mathcal{M}_1} \tau_j^* + a^* =1$ holds at the optimal solution. The following Lemma establishes the relation among $\{a^*,\boldsymbol{\tau}^*,\nu^*\}$.

\textbf{Lemma $2$:} The optimal $\{a^*,\boldsymbol{\tau}^*,\nu^*\}$ satisfies
\begin{equation}
\label{7}
\frac{\tau_j^*}{a^*} = \frac{\eta_2 h_j^2}{-\left(W\left(-\frac{1}{\exp(1+\frac{\nu^*}{w_j\varepsilon})}\right)\right)^{-1}-1}, \ \forall j\in \mathcal{M}_1,
\end{equation}
where $W(x)$ denotes the Lambert-W function, which is the inverse function of $f(z) = z \exp(z) =x$, i.e., $z = W(x)$.

\emph{Proof:} Please see the detailed proof in the Appendix A. $\hfill \blacksquare$

We can infer from (\ref{7}) that $\nu^*> 0$ holds strictly, because otherwise either $\tau_j^* \rightarrow \infty$ or $a^*\rightarrow 0$ must hold, which are evidently not true at optimum. When $\nu^*> 0$, we have $-1/e < - \frac{1}{\exp(1+\frac{\nu^*}{w_j\varepsilon}) }< 0$. As $W(x)\in(-1,0)$ when $x\in (-1/e,0)$, the denominator of the RHS of (\ref{7}) is always positive. Meanwhile, because $W(x)$ is an increasing function when $x\in (-1/e,0)$, we can infer that a longer offloading time $\tau_j^*$ is allocated to WD with stronger wireless channels (larger $h_j$) and larger weight $w_j$. Let us denote (\ref{7}) as
\begin{equation}
\label{10}
\tau_j^* = \eta_2 h_j^2 a^* \cdot \varphi_j(\nu^*), \ \forall j\in \mathcal{M}_1,
\end{equation}
where
\begin{equation}
\label{142}
\varphi_j(\nu) \triangleq \left[-\left(W\left(-\frac{1}{\exp(1+\frac{\nu}{w_j\varepsilon })}\right)\right)^{-1}-1\right]^{-1}
\end{equation}
is a decreasing function in $\nu$, with $\varphi_j(\nu) \rightarrow \infty$ when $\nu\rightarrow 0$, and $\varphi_j(\nu) \rightarrow 0$ when $\nu\rightarrow \infty$.

By substituting (\ref{10}) into $\sum_{j\in \mathcal{M}_1} \tau_j^* + a^* = 1$, we obtain a semi-closed-form of $a^*$ as a function of $\nu^*$
\begin{equation}
\label{8}
\begin{aligned}
a^* = \frac{1}{1+ \eta_2 \left(\sum_{j\in \mathcal{M}_1} h_j^2\varphi_j(\nu^*)\right)} \triangleq p_1(\nu^*).
\end{aligned}
\end{equation}
Given the monotonicity of $\varphi_j(\nu^*)$, we can infer that $p_1(\nu)$ is an increasing function in $\nu$. In particular, $p_1(\nu)\rightarrow 0$ when $\nu \rightarrow 0$, and $p_1(\nu)\rightarrow 1$ when $\nu \rightarrow \infty$. We then have the following Proposition $1$ on the optimal value of $\nu^*$.

\textbf{Proposition 1}: There exists a unique optimal $\nu^*$ that satisfies
\begin{equation}
\label{19}
Q(\nu^*) \triangleq \frac{1}{3} \left(p_1(\nu^*)\right)^{-\frac{2}{3}}\sum_{i \in \mathcal{M}_0} w_i  \eta_1 \left(\frac{h_i}{k_i}\right)^{\frac{1}{3}} + \varepsilon \eta_2 \sum_{j\in \mathcal{M}_1} \frac{w_j h_j^2}{1+1/\varphi_j(\nu^*)} - \nu^* = 0,
\end{equation}
where $Q(\nu)$ is a monotonically decreasing function in $\nu >0$.

\emph{Proof:} Please see the detailed proof in the Appendix B. $\hfill \blacksquare$

With Proposition $1$, the optimal $\nu^*$ can be efficiently obtained via a bi-section search over $\nu \in (0, \bar{\nu})$ to find the unique $\nu$ that satisfies $Q(\nu)=0$, where $\bar{\nu}$ is a sufficiently large value. Now that the optimal $\nu^*$ is obtained, the optimal $\{a^*,\boldsymbol{\tau}^*\}$ can be directly calculated using (\ref{10}) and (\ref{8}). Due to the convexity, the primal and dual optimal values are the same for (P2) given $\mathcal{M}_0$. The pseudo-code of the bi-section search method is illustrated in Algorithm $1$. Given a precision parameter $\sigma_0$, it takes $O\left(\log_2\left(\frac{\bar{\nu}}{\sigma_0}\right)\right)$ number iterations for Algorithm $1$ to converge. In each iteration, the computational complexity of evaluating $Q(\nu)$ is proportional to the number of WDs, i.e., $O(N)$. Therefore, the overall complexity of Algorithm $1$ is $O(N)$. Compared with conventional interior point method with $O\left(N^3\right)$ complexity \cite{2004:Boyd}, the proposed algorithm significantly reduces the computational cost especially for large $N$. Besides, the calculation of the proposed algorithm involves only basic function evaluations, which is much easier to implement in hardware-constrained IoT networks than generic convex optimization algorithms.

\begin{algorithm}
\small
 \SetAlgoLined
 \SetKwData{Left}{left}\SetKwData{This}{this}\SetKwData{Up}{up}
 \SetKwRepeat{doWhile}{do}{while}
 \SetKwFunction{Union}{Union}\SetKwFunction{FindCompress}{FindCompress}
 \SetKwInOut{Input}{input}\SetKwInOut{Output}{output}
  \Input{WD mode selection $\{\mathcal{M}_0,\mathcal{M}_1\}$}
 \Output{the optimal $\{a^*,\boldsymbol{\tau}^*\}$ to Problem (P2) given $\mathcal{M}_0$}
 \textbf{initialization:}  $\sigma_0 \leftarrow 0.005$, $\bar{\nu}\leftarrow$ sufficiently large value\;
 $UB \leftarrow \bar{\nu}$, $LB \leftarrow 0$\;
       \Repeat{$|UB-LB|\leq \sigma_0$}{
      $\nu \leftarrow \frac{UB+LB}{2}$\;
      \eIf{$Q(\nu)>0$ in the LHS of (\ref{19})}{
      $LB \leftarrow \nu$\;
      }{
      $UB \leftarrow \nu$\;
      }
 }
  Calculate $a^*$ using (\ref{8}), and $\boldsymbol{\tau}^*$ using $a^*$ and (\ref{10})\;
 \textbf{Return} $\{a^*,\boldsymbol{\tau}^*\}$\;
 \caption{Bi-section search algorithm for optimal transmission time allocation}
\end{algorithm}

\subsection{Coordinate Descent Method for Computing Mode Optimization}
In this subsection, we propose a simple CD method to optimize $\mathcal{M}_0$. To facilitate the illustration, we introduce $N$ auxiliary binary variables $\mathbf{m} = [m_1,\cdots,m_N]'$, where $m_i=0$ (or $m_i=1$) denotes that a WD $i\in \mathcal{M}_0$ (or $i\in \mathcal{M}_1$). Because each $\mathbf{m}$ corresponds to a unique mode selection solution $\mathcal{M}_0$ in (P2), it is equivalent to optimize $\mathbf{m}$ for solving (P2).

Now that jointly optimizing the $N$ binary variables is difficult, the CD method successively optimizes along the direction of only one variable $m_i$ (i.e., the coordinate direction) at a time to find the local maximum \cite{2009:Rao}. Specifically, starting with an initial $\mathbf{m}^0$, we denote $\mathbf{m}^{l-1}$ as the mode selection decision at the $(l-1)$-th iteration, $l=1,2,\cdots$. Correspondingly, we denote $V\left(\mathbf{m}^{l-1}\right)$ as the optimal value of (P2) given $\mathbf{m}^{l-1}$, which can be obtained using Algorithm $1$. Let $R_j^{l}$ denote the reward if WD$_j$ swaps its current computing mode in the $l$-th iteration, defined as the increase of objective value of (P2) after the swapping, i.e.,
\begin{equation}
\label{36}
R_j^l =  V\left(\mathbf{m}^{l-1}(j)\right) - V\left(\mathbf{m}^{l-1}\right),
\end{equation}
where $\mathbf{m}^{l-1}(j)$ denotes the mode selection after WD$_j$ swaps its current mode, i.e.,
\begin{equation}
\label{117}
\mathbf{m}^{l-1}(j) = \left[m_1^{l-1},m_2^{l-1}\cdots, m_j^{l-1}\oplus1,\cdots,m_{N-1}^{l-1},m_N^{l-1}\right]'.
\end{equation}
Here, $\oplus$ denotes the modulo-2 summation operator, e.g., $1\oplus1 = 0$. Then, we obtain the mode selection in the $l$-th iteration, $\mathbf{m}^{l}$, by letting the WD that achieves the highest reward swap its computing mode, if the reward is positive. In other words, $\mathbf{m}^{l} = \mathbf{m}^{l-1}(j^*_l)$ if $R_{j^*_l}^l>0$, where $j^*_l = \arg \max_{j=1,\cdots,N} R_j^l$. The pseudo-code of the method is illustrated in Algorithm $2$. The objective function value of (P2) increases monotonically as the iterations proceed. Meanwhile, the optimal value of (P2) is bounded, thus the CD method guarantees to converge. Nonetheless, the convergence speed could be slow in large-size networks with high searching dimensions.

\begin{algorithm}
\small
 \SetAlgoLined
 \SetKwData{Left}{left}\SetKwData{This}{this}\SetKwData{Up}{up}
 \SetKwRepeat{doWhile}{do}{while}
 \SetKwFunction{Union}{Union}\SetKwFunction{FindCompress}{FindCompress}
 \SetKwInOut{Input}{input}\SetKwInOut{Output}{output}
 \Input{Initial mode selection $\mathbf{m}^{0}$}
 \Output{An approximate solution $\{\bar{a},\boldsymbol{\bar{\tau}}, \bar{\mathbf{\mathcal{M}}}_0\}$ to (P2)}
 \textbf{initialization:}  $l\leftarrow 0$\;
       \Repeat{$v_l^* \leq 0$}{
       $l \leftarrow l+1$\;
      \For{\emph{each WD$_j$}}{
      Calculate $R_j^l$ in (\ref{36}) using Algorithm $1$\;
      }
      $v_l^*\leftarrow \max_{j=1,\cdots,N} R_j^l$ and $j^*_{l} \leftarrow \arg \max_{j=1,\cdots,N} R_j^l$\;
      Update $\mathbf{m}^{l} \leftarrow \mathbf{m}^{l-1}(j^*_{l})$ using (\ref{117})\;
 }
 Find the corresponding $\bar{\mathbf{\mathcal{M}}}_0$ given $\mathbf{m}^{l-1}$, $\left\{\bar{a},\boldsymbol{\bar{\tau}}\right\}\leftarrow$ the optimal solution of (P2) given  $\bar{\mathbf{\mathcal{M}}}_0$\;
 \textbf{Return} An approximate solution $\{\bar{a},\boldsymbol{\bar{\tau}}, \bar{\mathbf{\mathcal{M}}}_0\}$ to (P2)\;
 \caption{Coordinate descent algorithm for mode selection optimization.}
\end{algorithm}

\subsection{A Homogeneous Special Case}
In this subsection, we derive some interesting design insights from studying a special case with homogeneous WDs, where the weight and computation energy efficiency, i.e., $w_i=w$ and $k_i=k$, are equal for all the WDs. In this case, the WDs differ only by the wireless channel gain $h_i$'s. For those mode-$1$ WDs, it holds that $\varphi_j(\nu^*)$'s in (\ref{142}) are equal at optimum given the same $w_j=w$. Accordingly, we denote $\varphi_j(\nu^*) = \varphi(\nu^*)$, $\forall j \in \mathcal{M}_1$, and express the optimal computation rate of a mode-$1$ WD$_j$ by substituting (\ref{10}) to (\ref{4}), where
\begin{equation}
\label{143}
r_{O,j}^* = h_j^2 \cdot \varepsilon \eta_2 a \varphi(\nu^*) \ln\left(1+ \frac{1}{\varphi(\nu^*)}\right),\ \forall j\in \mathcal{M}_1.
\end{equation}
Because $\varphi(\nu)$ is a decreasing function and $\varphi\ln(1+\frac{1}{\varphi})$ increases with $\varphi>0$, we can infer that $r_{O,j}^*$ decreases with $\nu^*$. Intuitively, $\nu^*$ can be considered as the ``price" of the offloading time charged to the mode-$1$ WDs, which reflects the level of competitions in data offloading, e.g., the number and the channel conditions of offloading WDs. Besides, we can infer from (\ref{143}) that the mode-$1$ WDs offload to the AP at the \emph{same spectral efficiency}, but with different durations that are proportional to the square of wireless channel gain (indeed the product of uplink and downlink channel gains). Therefore, the computation rates are proportional to $h_j^2$'s as well. Intuitively, this is caused by both channel-related energy harvesting in the downlink and task offloading in the uplink. Then, a mode-$1$ WD with relatively weak channel (say $1/10$ of another mode-$1$ WD) may have much lower computation rate than the other mode-$1$ WDs ($1/100$ in this case).

On the other hand, the local computation rate of a mode-$0$ WD is only related to its own channel gain $h_i$, while irrespective to the other WDs' computing modes and channel conditions. Meanwhile, the computation rate $r_{L,i}^*$ decays slowly as $h_i$ decreases, i.e., $r_{L,i}^* \propto h_i^{\frac{1}{3}}$. For instance, a $10$ times stronger channel translates to only $2.15$ times higher computation rate. We can infer from the above analysis that the computation rate of a mode-$1$ WD is more sensitive to the wireless channel condition than a mode-$0$ WD. Intuitively, this indicates that a WD with relatively weak channel is likely to operate in local computing mode at the optimum of (P2), because otherwise operating in offloading mode may result in very small offloading time allocated to it, and thus significantly low computation rate, and vice versa. Interestingly, we have observed in the simulation section that the optimal computing mode selection of homogeneous WDs has a threshold structure based on the wireless channel gains. That is, the optimal mode selection solution $\left\{\mathcal{M}_0^*,\mathcal{M}_1^*\right\}$ to (P2) satisfies $h_j\geq h_i$, $\forall j \in \mathcal{M}_1^*$ and $\forall i\in \mathcal{M}_0^*$. In other words, the mode-$1$ WDs have stronger wireless channels than the other mode-$0$ WDs at the optimum.

\section{Joint Optimization using ADMM-Based Method}\label{joint}
The major advantage of the CD method proposed in the last section is its simplicity in implementation, because the computation involves basic function evaluations only. However, the local searching nature makes the CD method susceptible to high computational complexity in large-size networks with high searching dimensions. To address the problem in large-size networks with tens to several hundred of WDs, we propose in this section an ADMM-based algorithm to jointly optimize the computing mode selection and transmission time allocation. As we will show later, the proposed ADMM-based approach has a computational complexity that increases slowly with the network size $N$.

The main idea is to decompose the hard combinatorial optimization (P2) into $N$ parallel smaller integer programming problems, one for each WD. Nonetheless, conventional decomposition techniques, such as dual decomposition, cannot be directly applied to (P2) due to the coupling variable $a$ and constraint (\ref{116}) among the WDs. To eliminate these coupling factors, we first reformulate (P2) as an equivalent integer programming problem by introducing binary decision variables $m_i$'s and additional artificial variables $x_i$'s and $z_i$'s as follows
 \begin{subequations}
\label{12}
   \begin{align}
    (P3):\ \ & \underset{a, \mathbf{z},\mathbf{x},\boldsymbol{\tau},\mathbf{m}}{\text{maximize}} & &  \sum_{i=1}^N w_i \left\{ \left(1-m_i\right)\eta_1 \left(\frac{h_i}{k_i}\right)^{\frac{1}{3}} x_i^{\frac{1}{3}} +  m_i \varepsilon \tau_i \ln\left(1+ \frac{\eta_2 h_i^2 x_i}{\tau_i}\right)\right\}\\
    & \text{subject to} &  & \sum_{i=1}^N z_i + a \leq 1,\\
    &  & & x_i = a, z_i = \tau_i \ i=1,\cdots,N, \\
    &  & &   a, z_i, x_i,\tau_i\geq 0,\ m_i\in\left\{0,1\right\}, \ i=1,\cdots,N.
   \end{align}
\end{subequations}
Here, $m_i=0$ for all $i\in \mathcal{M}_0$ and $m_i=1$ for all $i\in\mathcal{M}_1$. $\mathbf{z}=[z_1,\cdots,z_N]'$ and $\mathbf{x} = [x_1,\cdots,x_N]'$. With a bit abuse of notation, we denote $\boldsymbol{\tau} = [\tau_1,\cdots,\tau_N]'$. Notice that variables $z_i$ and $\tau_i$ are immaterial to the objective if $m_i = 0$. Then, (P3) can be equivalently written as
\begin{subequations}
\label{13}
   \begin{align}
    & \underset{a, \mathbf{z},\mathbf{x},\boldsymbol{\tau},\mathbf{m}}{\text{maximize}} & &  \sum_{i=1}^N  q_i(x_i,\tau_i,m_i) + g(\mathbf{z},a)\\
    & \text{subject to} &  & x_i = a, \tau_i = z_i \ i=1,\cdots,N, \label{122}\\
    &  & &   x_i,\tau_i\geq 0,\ m_i\in\left\{0,1\right\}, \ i=1,\cdots,N,
   \end{align}
\end{subequations}
where
\begin{equation}
q_i(x_i,\tau_i,m_i) = w_i \left\{ \left(1-m_i\right)\eta_1 \left(\frac{h_i}{k_i}\right)^{\frac{1}{3}} x_i^{\frac{1}{3}} +  m_i \varepsilon \tau_i \ln\left(1+ \frac{\eta_2 h_i^2 x_i}{\tau_i}\right)\right\},
\end{equation}
and
\begin{equation}
\label{42}
g(\mathbf{z},a) = \begin{cases}
0,&   \text{if} \left(\mathbf{z},a\right) \in \mathcal{G},\\
-\infty,   &   \text{otherwise},\\
\end{cases}
\end{equation}
where $\mathcal{G} = \left\{\left(\mathbf{z},a\right)\mid \sum_{i=1}^N z_i +a \leq 1, a \geq 0, z_i \geq 0, i=1,\cdots,N \right\}$.

Problem (\ref{13}) can be effectively decomposed using the ADMM technique \cite{2011:Boyd}, which solves for the optimal solution of the dual problem. By introducing multipliers to the constraints in (\ref{122}), we can write a partial augmented Lagrangian of (\ref{13}) as
\begin{equation}
\begin{aligned}
L\left(\mathbf{u},\mathbf{v},\boldsymbol{ \theta}\right) =&  \sum_{i=1}^N  q_i(\mathbf{u}) + g(\mathbf{v}) + \sum_{i=1}^N \beta_i \left(x_i - a\right) + \sum_{i=1}^N \gamma_i \left(\tau_i - z_i\right)  \\
&- \frac{c}{2} \sum_{i=1}^N \left(x_i - a\right)^2 - \frac{c}{2} \sum_{i=1}^N \left(\tau_i - z_i\right)^2,
\end{aligned}
\end{equation}
where $\mathbf{u} = \left\{\mathbf{x},\boldsymbol{\tau},\mathbf{m}\right\}$, $\mathbf{v}=\left\{\mathbf{z},a\right\}$, and $\boldsymbol{\theta} = \{\boldsymbol{\beta},\boldsymbol{\gamma}\}$. $c>0$ is a fixed step size. The corresponding dual function is
\begin{equation}
d(\boldsymbol{\theta}) = \underset{\mathbf{u},\mathbf{v}}{\text{maximize}}\ \left\{L\left(\mathbf{u},\mathbf{v},\boldsymbol{ \theta}\right)\mid \mathbf{x}\geq \mathbf{0},\boldsymbol{\tau}\geq \mathbf{0},\mathbf{m}\in \mathbb{B}^{N\times 1}\right\},
\end{equation}
where $\mathbb{B}^{N\times 1}$ denotes a $(N\times 1)$ binary vector. Furthermore, the dual problem is
\begin{equation}
\label{37}
\underset{\boldsymbol{\theta}}{\text{minimize}}\ d\left(\boldsymbol{\theta}\right).
\end{equation}

The ADMM technique solves the dual problem (\ref{37}) by iteratively updating $\mathbf{u}$, $\mathbf{v}$, and $\boldsymbol{\theta}$. We denote the values in the $l$-th iteration as $\left\{\mathbf{u}^l, \mathbf{v}^l, \boldsymbol{\theta}^l\right\}$. Then, in the $(l+1)$-th iteration, the update of the variables is performed sequentially as follows:
\subsubsection{Step 1} Given $\left\{\mathbf{v}^l,\boldsymbol{\theta}^l\right\}$, we first maximize $L$ with respect to $\mathbf{u}$, where
\begin{equation}
\label{20}
\mathbf{u}^{l+1} = \arg \ \underset{\mathbf{u}}{\text{maximize}}\  L\left(\mathbf{u},\mathbf{v}^l,\boldsymbol{\theta}^l\right).
\end{equation}
Notice that (\ref{20}) can be decomposed into $N$ parallel subproblems. Each subproblem solves
\begin{equation}
\label{38}
\{x_i^{l+1},\tau_i^{l+1},m_i^{l+1}\}= \arg \underset{x_i, \tau_i \geq 0, m_i\in\{0,1\}}{\text{maximize}}\ s^l(x_i,\tau_i,m_i),
\end{equation}
where
\begin{equation}
s_i^l(x_i,\tau_i,m_i) = q_i\left(x_i,\tau_i,m_i\right) + \beta_i^{l} x_i + \gamma_i^l \tau_i - \frac{c}{2}\left(x_i-a^l\right)^2 - \frac{c}{2}\left(\tau_i - z_i^l\right)^2.
\end{equation}

By considering $m_i=0$ and $1$, respectively, we can express (\ref{38}) as
\begin{equation}
\label{16}
\begin{aligned}
\begin{cases}
\underset{x_i,\tau_i\geq 0}{\text{maximize}}\  w_i \eta_1 \left(\frac{h_i}{k_i}\right)^{\frac{1}{3}} x_i^{\frac{1}{3}}+ \beta_i^{l} x_i + \gamma_i^l \tau_i - \frac{c}{2}\left(x_i-a^l\right)^2 - \frac{c}{2}\left(\tau_i - z_i^l\right)^2,  & m_i=0,\\
\underset{x_i,\tau_i\geq 0}{\text{maximize}}\ w_i \varepsilon \tau_i \ln \left(1+\frac{\eta_2 h_i^2 x_i}{ \tau_i}\right)+ \beta_i^{l} x_i + \gamma_i^l \tau_i - \frac{c}{2}\left(x_i-a^l\right)^2 - \frac{c}{2}\left(\tau_i - z_i^l\right)^2, & m_i=1.
\end{cases}
\end{aligned}
\end{equation}
For both $m_i=0$ and $1$, (\ref{16}) solves a strictly convex problem, and thus the optimal solution can be easily obtained, e.g., using the projected Newton's method \cite{2004:Boyd}. Accordingly, we can simply select $m_i=0$ or $1$ that yields a larger objective value in (\ref{16}) as $m_i^{l+1}$, and the corresponding optimal solution as $x_i^{l+1}$ and $\tau_i^{l+1}$. After solving the $N$ parallel subproblems, the optimal solution to (\ref{20}) is given by $\mathbf{u}^{l+1} = \left\{\mathbf{x}^{l+1},\boldsymbol{\tau}^{l+1},\mathbf{m}^{l+1}\right\}$. Notice that the complexity of solving each subproblem in (\ref{16}) does not scale with $N$ (i.e., $O(1)$ complexity), thus the overall computational complexity of Step $1$ is $O(N)$.

\subsubsection{Step 2} Given $\mathbf{u}^{l+1}$, we then maximize $L$ with respect to $\mathbf{v}$. By the definition of $g(\mathbf{v})$ in (\ref{42}), $\mathbf{v}^{l+1} \in \mathcal{G}$ must hold at the optimum. Accordingly, the maximization problem can be equivalently written as the following convex optimization problem
 \begin{subequations}
 \label{43}
\begin{align}
\mathbf{v}^{l+1} = &\arg \underset{ \mathbf{z},a}{\text{maximize}} & & \sum_{i=1}^N \beta_i^l \left(x_i^{l+1} - a\right) + \sum_{i=1}^N \gamma_i^l \left(\tau_i^{l+1} - z_i\right) \\
 & & &- \frac{c}{2} \sum_{i=1}^N \left(x_i^{l+1} - a\right)^2 - \frac{c}{2} \sum_{i=1}^N \left(\tau_i^{l+1} - z_i\right)^2\\
 & \text{subject to} &  & \sum_{i=1}^N z_i +a \leq 1,\ a\geq 0, \ z_i\geq 0, i=1,\cdots,N.
\end{align}
 \end{subequations}
Instead of using standard convex optimization algorithms to solve (\ref{43}), e.g., interior point method, here we devise an alternative low-complexity algorithm. By introducing a multiplier $\psi$ to the constraint $\sum_{i=1}^N z_i +a \leq 1$, it holds at the optimum that
\begin{equation}
\begin{aligned}
a^* &= \left(\frac{\sum_{i=1}^N x_i^{l+1}}{N} - \frac{\sum_{i=1}^N \beta_i^l + \psi^*}{cN}\right)^{+},\\
z_i^* &= \left(\tau_i^{l+1} - \frac{\gamma_i^l + \psi^*}{c}\right)^{+},\ i=1,\cdots,N,
\end{aligned}
\end{equation}
where $(x)^+ \triangleq \max\left(x,0\right)$. As $a^*$ and $z_i^*$ are non-increasing with $\psi^* \geq 0$, the optimal solution can be obtained by a bi-section search over $\psi^*\in (0,\bar{\psi})$, where $\bar{\psi}$ is a sufficiently large value, until $\sum_{i=1}^N z_i^* +a^* = 1$ is satisfied (if possible), and then comparing the result with the case of $\psi^* =0$ (the case that $\sum_{i=1}^N z_i^* +a^* < 1$). The details are omitted due to the page limit. Overall, the computational complexity of the bi-section search method to solve (\ref{43}) is $O(N)$.

\subsubsection{Step 3} Finally, given $\mathbf{u}^{l+1}$ and $\mathbf{v}^{l+1}$, we minimize $L$ with respect to $\boldsymbol{\theta}$, which is achieved by updating the multipliers $\boldsymbol{\theta}^l= \{\boldsymbol{\beta}^{l},\boldsymbol{\gamma}^l\}$ as
\begin{equation}
\label{22}
\begin{aligned}
\beta_i^{l+1} &= \beta_i^l - c(x_i^{l+1}-a^{l+1}),\ i=1,\cdots,N,\\
\gamma_i^{l+1} & = \gamma_i^l - c(\tau_i^{l+1}-z_i^{l+1}),\ i=1,\cdots,N.\\
\end{aligned}
\end{equation}
Evidently, the computational complexity of Step $3$ is $O(N)$ as well.

\begin{algorithm}
\small
 \SetAlgoLined
 \SetKwData{Left}{left}\SetKwData{This}{this}\SetKwData{Up}{up}
 \SetKwRepeat{doWhile}{do}{while}
 \SetKwFunction{Union}{Union}\SetKwFunction{FindCompress}{FindCompress}
 \SetKwInOut{Input}{input}\SetKwInOut{Output}{output}
 \Input{The number of WDs $N$ and other system parameters, e.g, $h_i$'s and $w_i$'s.}
 \textbf{initialization:}  $\{\boldsymbol{\beta}^0,\boldsymbol{\gamma}^0\}\leftarrow -100$;  $a^0 \leftarrow 0.9$;  $z_i^0 = (1-a^0)/N,\ i=1,\cdots, N$\;
 $c\leftarrow \varepsilon$, $\sigma_1\leftarrow 0.0005N$ , $l\leftarrow 0$\;
       \Repeat{$\sum_{i=1}^N \left(|x_i^{l}-a^{l}|+ |\tau_i^l-z^l|\right)<2\sigma_1$ \emph{and} $|a^{l}-a^{l-1}|+\sum_{i=1}^N |z_i^{l}-z_i^{l-1}|<\sigma_1$}{
      \For{\emph{each} WD$_i$}{
            Update local variables $\{x_i^{l+1},\tau_i^{l+1}, m_i^{l+1}\}$ by solving (\ref{16})\;
      }
      Update coupling variables $\left\{\mathbf{z}^{l+1},a^{l+1}\right\}$ by solving (\ref{43})\;

      Update multipliers $\{\boldsymbol{\beta}^{l+1},\boldsymbol{\gamma}^{l+1}\}$ using (\ref{22})\;
      $l\leftarrow l+1$\;
}
 \textbf{Return} $\left\{a^l,\boldsymbol{\tau}^l,\mathbf{m}^l\right\}$ as an approximate solution to (P3)\;
 \caption{ADMM-based joint mode selection and resource allocation algorithm}
\end{algorithm}

The above Steps $1$ to $3$ repeat until a specified stopping criterion is met. In general, the stopping criterion is specified by two thresholds: absolute tolerance (e.g., $\sum_{i=1}^N |x_i^{l}-a^{l}|+ |\tau_i^l-z^l|$) and relative tolerance (e.g., $|a^{l}-a^{l-1}|+\sum_{i=1}^N |z_i^{l}-z_i^{l-1}|$) \cite{2011:Boyd}. The pseudo-code of the ADMM method solving (P2) is illustrated in Algorithm $3$. As the dual problem (\ref{37}) is convex in $\boldsymbol{\theta}= \{\boldsymbol{\beta},\boldsymbol{\gamma}\}$, the convergence of the proposed method is guaranteed. Meanwhile, the convergence of the ADMM method is insensitive to the choice of step size $c$ \cite{2015:Ghadimi}. Thus, we set $c = \varepsilon$ without loss of generality. Besides, we can infer that the computational complexity of one ADMM iteration (including the $3$ steps) is $O(N)$, because each of the $3$ steps has $O(N)$ complexity. Notice that the ADMM algorithm may not exactly converge to the primal optimal solution of (P3) due to the potential duality gap of non-convex problems. Therefore, upon termination of the algorithm, the dual optimal solution $\left\{a^l,\boldsymbol{\tau}^l,\mathbf{m}^l\right\}$ is an approximate solution to (P3), whose performance gap will be evaluated through simulations.

\section{Extensions and Discussions}
In this section, we discuss some potential extensions of the proposed CD and ADMM methods in other practical setups. For simplicity of illustration, we assume in this paper that the RF energy transmitter and edge server are integrated into a single AP with equal uplink (for computation offloading) and downlink (for WET) wireless channels. Nonetheless, the proposed CD and ADMM methods can be easily extended to the case with non-equal uplink channel $g_i$ and downlink channel $h_i$ without modifying the structure of the algorithms. This is achievable by simply replacing $h_i^2$ with $h_ig_i$ in the second term of the objective in (P2). In this sense, the proposed methods can also be used in a wireless powered MEC system where the RF energy transmitter and edge server are installed at two separate nodes. Besides, our proposed methods can be applied to solve the max-min rate optimization problem, which is a common formulation in wireless communication systems to enhance the \emph{user fairness} (e.g., see \cite{2014:Ju}). In our problem, the max-min formulation maximizes the minimum computation rate among the WDs. Our methods are applicable because a max-min rate optimization problem has its dual problem in the form of weighted-sum-rate-maximization like (P2). In this sense, the proposed methods can be applied to both enhance the computation efficiency and user fairness in a wireless powered MEC system.

Besides the proposed CD and ADMM methods, the technique of linear relaxation (LR) can also be applied to solve (P2) for an approximate solution. Specifically, we allow each WD$_i$ to arbitrarily partition its harvested energy $E_i$ for performing both local computation and computation offloading, denoted by $E_{L,i}$ and $E_{O,i}$, respectively. This is commonly referred to as the partial offloading policy. In this sense, the considered binary offloading policy corresponds to the case that either $E_{O,i}=0$ or $E_{O,i}=E_i$. Due to the page limit, we omit some details on formulation and express the linearly relaxed computation rate maximization problem as
\begin{subequations}
   \label{335}
   \begin{align}
    & \underset{\mathbf{e},a,\mathbf{f}, \boldsymbol{\tau}}{\text{maximize}} & &  \sum_{i=1}^N w_i  \left\{ \frac{f_i}{\phi} + \frac{B\tau_i}{v_u} \log_2\left(1 + \frac{e_{i} h_i}{\tau_i N_0}\right)\right\}\\
    & \text{subject to} &  & \sum_{i=1}^N \tau_i + a \leq 1,\\
    & & & k_if_i^3 + e_i \leq  \mu P h_i a,\ \ i=1,\cdots,N.\\
    & & & a \geq 0,\ \tau_i\geq 0, \ \ i=1,\cdots,N,
   \end{align}
\end{subequations}
where $\mathbf{e}\triangleq \left[e_1,\cdots,e_N\right]'$ and $e_i \triangleq \frac{E_{O,i}}{T}$. Notice that (\ref{335}) is a convex optimization problem that can be efficiently solved. Besides, its optimal objective value provides a performance upper bound to (P2). In general, both $E_{L,i}^*>0$ and $E_{O,i}^*>0$ hold at the optimum for some WD$_i$'s, indicating that these WDs perform both local computation and offloading. To find a feasible binary offloading solution to (P2), we can simply round the optimal solution of (\ref{335}), such that a WD chooses mode-$0$ if its local computation rate is higher than its offloading rate, and mode-$1$ otherwise. We refer to the method as LR-Round scheme. Then, the computation rate of the LR-Round scheme can be obtained by substituting the rounded computing modes of all the WDs in (P2), where the details are omitted. The upper bound achieved by the LR formulation in (\ref{335}) and the LR-Round scheme will be used as performance benchmarks in simulations.

\section{Simulation Results}
In this section, we present simulations to verify our analysis and evaluate the performance of the proposed algorithms. In all simulations, we use the parameters of the Powercast TX91501-3W transmitter with $P=3$W (Watt) as the energy transmitter at the AP, and those of P2110 Powerharvester as the energy receiver at each WD with $\mu= 0.51$ energy harvesting efficiency.\footnote{Please see the detailed product specifications on the website of Powercast Co. (http://www.powercastco.com).} Without loss of generality, we set $T=1$. Unless otherwise stated, we consider a Rayleigh fading channel model, where the channel gain $h_i = \bar{h}_i \alpha$. Here, $\bar{h}_i$ denotes the average channel gain determined by the location of the $i$-th WD and $\alpha$ denotes an independent exponential random variable of unit mean. Specifically, $\bar{h}_i$ follows the free-space path loss model
\begin{equation}
\bar{h}_i = A_d\left(\frac{3\cdot10^8}{4\pi f_c d_i}\right)^{d_e},\ i=1,\cdots,N,
\end{equation}
where $A_d = 4.11$ denotes the antenna gain, $f_c =915$ MHz denotes the carrier frequency, $d_i$ in meters denotes the distance between the WD$_i$ and AP, and $d_e\geq 2$ denotes the path loss exponent. Unless otherwise stated, $d_e=2.8$. Likewise, we set equal computing efficiency parameter $k_i = 10^{-26}$, $i=1,\cdots,N$, and $\phi=100$ for all the WDs \cite{2016:Wang}. For the data offloading mode, the bandwidth $B=2$ MHz and $v_u = 1.1$. In addition, the weighting factor $w_i$ is randomly assigned as either $1$ or $2$ with equal probability.

\subsection{Properties of Optimal Solution}
We first study some interesting properties of the optimal solution to (P2), which is obtained by enumerating all the $2^N$ combinations of the $N$ WDs' computing mode selections. For the simplicity of illustration, we consider $N=10$ and set $d_i = 2.5 + 0.3(i-1)$ meters, $i=1\cdots,10$. Besides, we consider a static channel model with $\alpha=1$ such that $h_i = \bar{h}_i$. In this case, the WDs are equally spaced by $0.3$ meters and the channel gain decreases from $h_1$ to $h_{10}$.

\begin{figure}
  \centering
  \subfigure[Optimal solution when $k_i = 10^{-26}$.]{\includegraphics[width=0.48\textwidth]{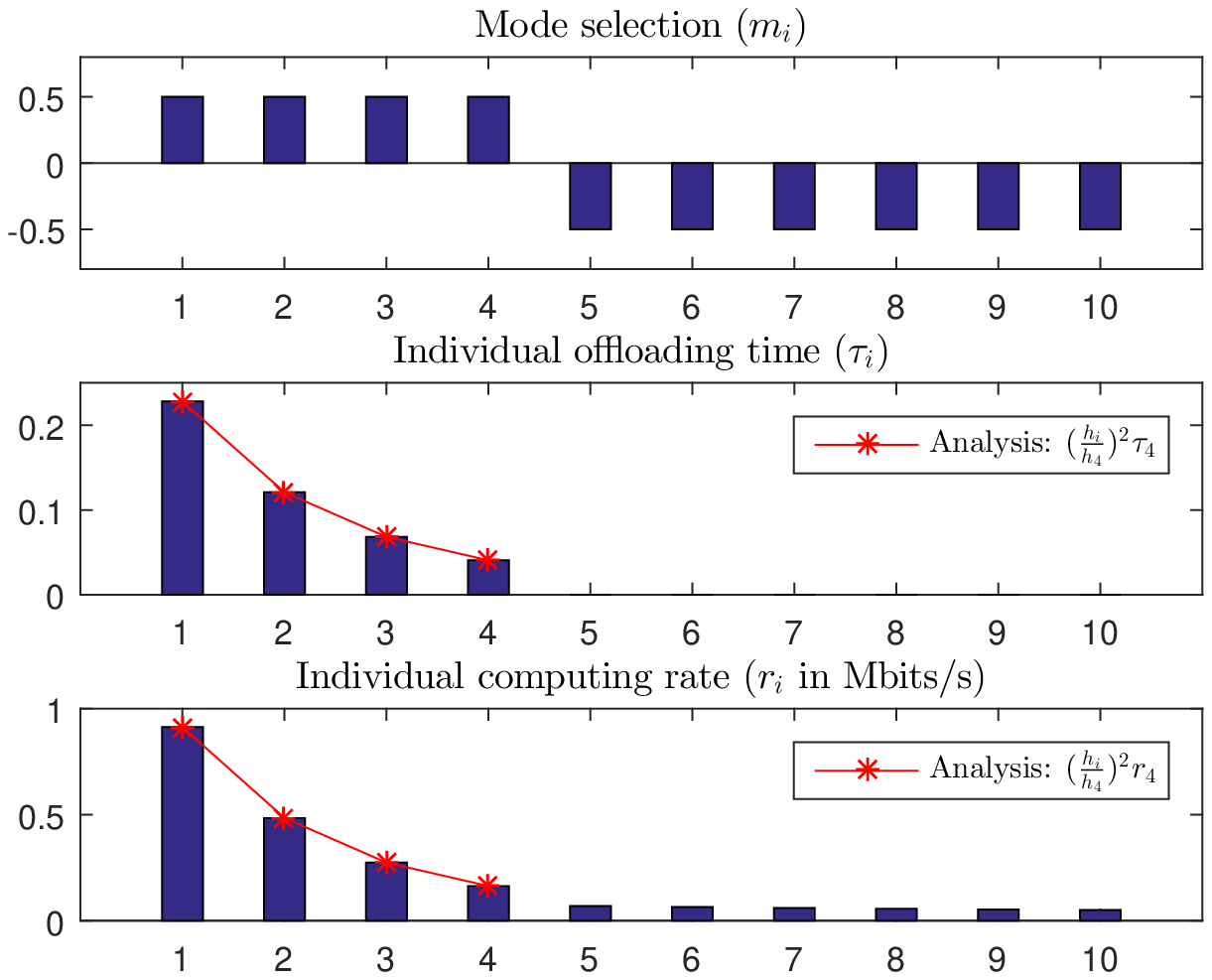}}\quad
  \subfigure[Optimal mode selection when $k_i$ varies.]{\includegraphics[width=0.48\textwidth]{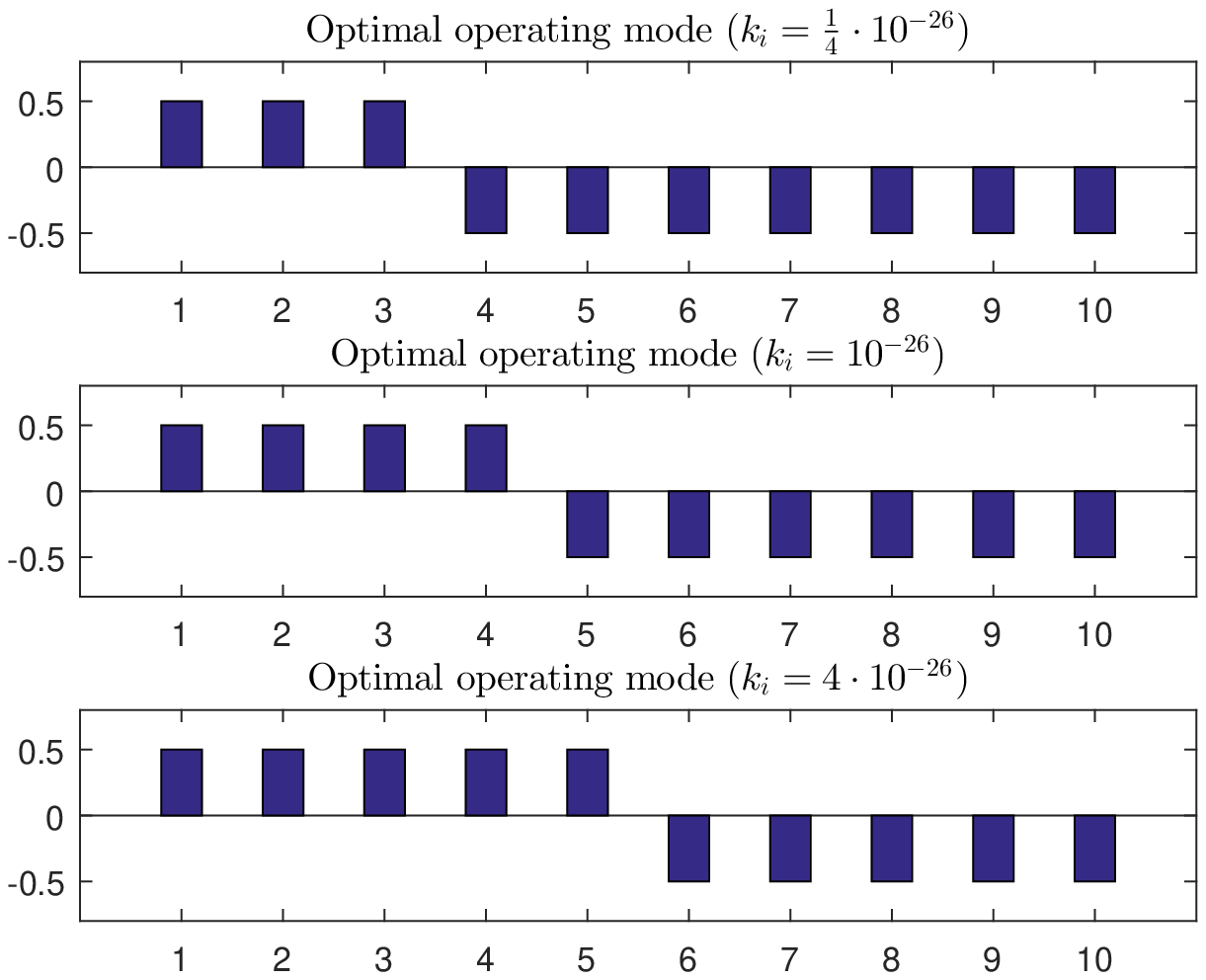}}
  \caption{Optimal solutions of the homogeneous special case of $10$ WDs with equal $k_i$ and $w_i$.}
  \label{102}
\end{figure}

In Fig.~\ref{102}, we first study a homogeneous special case with $w_i=1$ for all the WDs. In particular, we plot in Fig.~\ref{102}(a) the optimal mode selection (the figure above), the offloading time (the figure in the middle), and the individual computation rate (the bottom figure) of the $10$ WDs when computing efficiency $k_i = 10^{-26}$. In all the three sub-figures, the x-axis denotes the indices of the $10$ WDs. Without loss of generality, we use $m_i = 0.5$ and $m_i = -0.5$ to denote that a WD$_i$ selects mode $1$ and $0$, respectively. We can see that the optimal mode selection has a threshold structure, where the $4$ mode-$1$ WDs have stronger wireless channels than the other mode-$0$ WDs. Besides, both the optimal offloading time and the computation rates are proportional to $h_i^2$ for the mode-$1$ WDs, which matches with our analysis in Section IV.C. We also observe from the bottom figure of Fig.~\ref{102}(a) that the use of edge computing significantly improves the computation rate of the mode-$1$ WDs. In Fig.~\ref{102}(b), we further study the impact of computing efficiency $k_i$ to the optimal mode selection. From the top to the bottom figures, $k_i$ increases by $16$ times for all the WDs. Fewer WDs choose mode-$0$ as $k_i$ increases because local computation becomes less energy-efficient. Meanwhile, the optimal computing mode remains a threshold structure for all cases. In this sense, the optimal computing mode of a homogeneous special case can be easily obtained by searching the threshold from the WD with the strongest channel to the weakest WD. The theoretical proof of the threshold structure is left for future investigation.

\begin{figure}
\centering
  \begin{center}
    \includegraphics[width=0.55\textwidth]{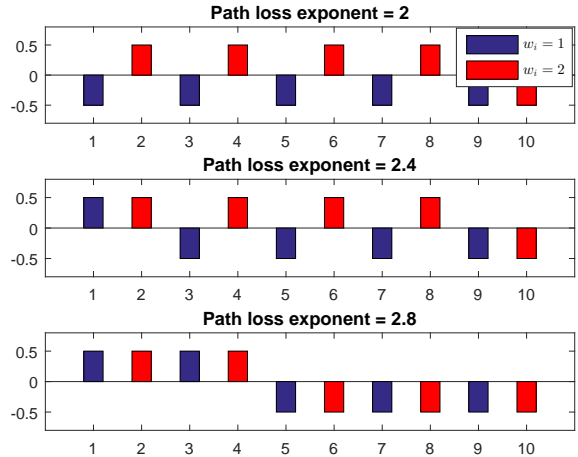}
  \end{center}
  \caption{Change of optimal computing modes of a heterogeneous case, where $w_i=1$ if $i$ is an odd number and $w_i=2$ otherwise. The three figures show the performance under path loss exponent $d_e=\{2,2.4,2.8\}$, respectively.}
  \label{103}
\end{figure}

In Fig.~\ref{103}, we consider a heterogeneous case, where the WDs have different weights $w_i$'s. For simplicity of illustration, we set $w_i=1$ if $i$ is an odd number and $w_i=2$ otherwise. We plot the variation of optimal computing modes when the path loss exponent $d_e \in \left\{2,2.4,2.8\right\}$. Notice that a larger $d_e$ indicates a larger channel disparity among the WDs and vice versa. When the wireless channel disparity is relatively small, the weighting factor plays an important rule in the mode selection. The four WDs with higher weights operate in mode $1$ when $d_e=2$. However, as the channel disparity increases, wireless channel condition becomes a more dominant factor. Now the four WDs with the strongest channels operate in mode $1$ when $d_e=2.8$. Interestingly, the optimal mode selection also has a threshold structure within each group of WDs with the equal weight. For instance, when $d_e=2.4$, for WDs with $w_i=1$, only the single WD with the strongest channel operates in mode $1$; while for WDs with $w_i=2$, the $4$ WDs with strongest channels operate in mode $1$.

\subsection{Computation Rate Performance Comparison}
In this subsection, we evaluate the computation rate performance of the proposed algorithms. For the CD method, the initial mode selection is randomly selected, while the initial condition of ADMM-based method is specified in Algorithm $3$. Besides, we consider the following three representative benchmark methods:
\begin{enumerate}
  \item Optimal: exhaustively enumerates all the $2^N$ combinations of $N$ WDs' computing modes and outputs the best performing one;
  \item Offloading only: all the WDs offload their tasks to the AP, $\mathcal{M}_0 = \emptyset$;
  \item Local computing only: all the WDs perform computations locally, $\mathcal{M}_0 = \mathcal{M}$.
\end{enumerate}

\begin{figure}
  \centering
  \subfigure[Under different path loss exponent.]{\includegraphics[width=0.48\textwidth]{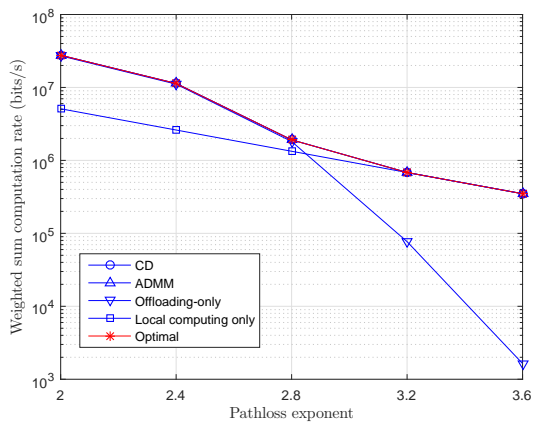}}\quad
  \subfigure[Under different average AP-to-WD distance.]{\includegraphics[width=0.48\textwidth]{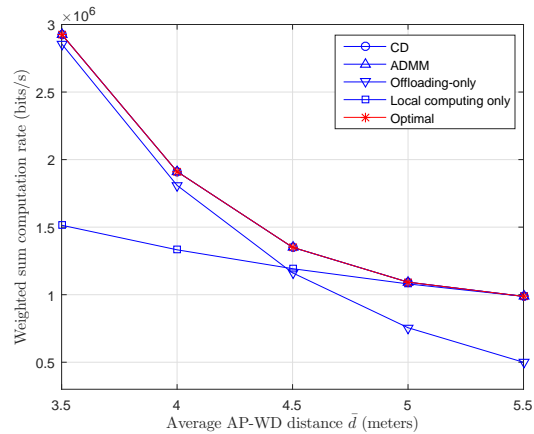}}
  \caption{Comparisons of computation rate performance of different algorithms. Left figure: when $d_e$ varies. Right figure: when the average AP-to-WD distance varies, the path-loss exponent is fixed as $d_e=2.8$.}
  \label{104}
\end{figure}

\begin{figure}
\centering
  \begin{center}
    \includegraphics[width=0.55\textwidth]{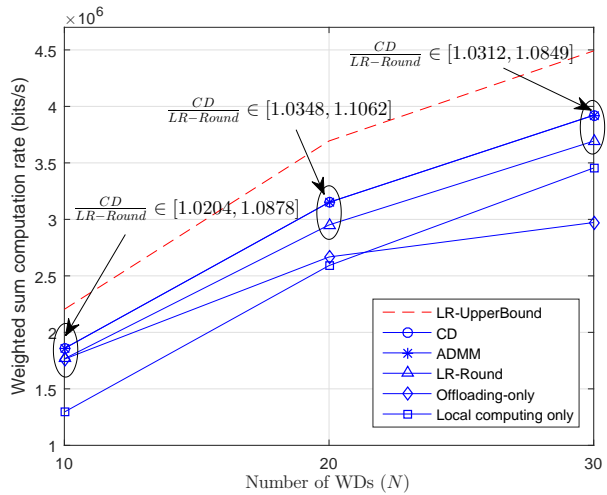}
  \end{center}
  \caption{Computation rate comparisons of different algorithms when the number of WDs varies.}
  \label{105}
\end{figure}

In Fig.~\ref{104} and \ref{105}, we compare the computation rate performance of different schemes under different network setups. Without loss of generality, we consider $10$ WDs, where each $d_i$ is independently generated from a truncated Gaussian distribution as $d_i = \min\left(\max\left(X,\bar{d}-1.5\right),\bar{d}+1.5\right)$, where $X \sim \mathcal{N}\left(\bar{d}, \sigma_d^2\right)$ is a Gaussian random variable with $\bar{d}$ denoting the average AP-to-WD distance and $\sigma_d$ denoting the standard deviation of placement spread. Each point in Fig.~\ref{104} and \ref{105} is an average performance of $20$ independent placements of the WDs, while the value of each placement is averaged over $100$ independent Rayleigh channel fading realizations.

In Fig.~\ref{104}(a), we set $\bar{d} = 4$ and $\sigma_d = 0.2$ and compare the computation rates when the path loss exponent $d_e$ increases from $2$ to $3.6$. We see that the proposed CD and ADMM methods both achieve near-optimal performance for all values of $d_e$ (at most $0.05\%$ performance gap compared to the optimal value), where the two curves are on top of each other with the optimal scheme. The offloading-only scheme can achieve close-to-optimal performance when $d_e$ is small such that the wireless channels are strong, but quickly degrades when $d_e$ increases, because the offloading rates severely suffer from the weak channels in both the uplink and downlink. The local local-computing-only scheme, on the other hand, achieves the worst performance when $d_e$ is small but near-optimal performance when $d_e\geq 3.2$. In Fig.~\ref{104}(b), we fix $d_e = 2.8$ and $\sigma_d = 0.2$ and compare the computation rates when the average AP-to-WD distance $\bar{d}$ varies. We observe that both the CD and ADMM methods achieve near-optimal performance for all values of $\bar{d}$. The offloading-only scheme achieves relatively good performance when $\bar{d}$ is small, e.g., $\bar{d}\leq 4$, but poor performance when $\bar{d}$ is large. The local-computing-only scheme, however, performs poorly when $\bar{d}$ is small but achieving near-optimal performance when $\bar{d}$ is large. The results in Fig.~\ref{104}(a) and (b) show that it is more preferable for a WD to offload computation when its wireless channel is strong and to perform local computing otherwise.

In Fig.~\ref{105}, we compare the performance of different algorithms when the number of WDs $N$ varies from $10$ to $30$. For each $N$, we assume that each $d_i$ follows the truncated Gaussian distribution with $\bar{d} =4$, $\sigma_d = 0.2$. The path-loss exponent is fixed as $d_e=2.8$. Because the optimal performance based on computing mode enumeration is computationally infeasible for $N>10$, we present here a performance upper bound obtained by linearly relaxing (LR) the binary offloading constraint. Besides, the LR-Round scheme is also considered for comparison.

In Fig.~\ref{105}, the proposed CD and ADMM methods have almost the identical performance, where the less than $0.05\%$ difference is mainly caused by the prescribed precision of convergence criterion. Besides, the CD and ADMM methods can achieve on average $86.3\%$ of the performance upper bound, despite that the bound is very loose in general. Meanwhile, there is an evident performance gap between the CD/ADMM method and the LR-Round scheme. On average, the computation rate of the CD/ADMM method is $6.3\%$ higher than the LR-Round scheme. Besides, we have also marked the range of the performance ratio (CD/LR-Round) in the figure for the $20\times 100=2000$ independent channel realizations. On one hand, we can see that the CD/ADMM method is strictly better than the LR-Round scheme in all the placement scenarios, i.e., the performance ratio is always larger than $1$. On the other hand, we can see that the LR-Round scheme is sensitive to the placement of the WDs. For instance, the computation rate of the LR-Round scheme is more than $10\%$ lower than the CD/ADMM method for some placement scenario when $N=20$. Intuitively, this is because the LR-Round scheme happens to wrongly select the computing mode of some WDs, where the resulted impact to the overall system performance is closely related to the location of all the WDs. In addition, we can also observe that the proposed CD/ADMM method significantly outperforms the other two benchmark methods, i.e., on average $18.5\%$ and $26.2\%$ higher than the offloading-only and local-computing-only schemes, respectively.

To sum up from Fig.~\ref{104} and \ref{105}, the performance of the considered benchmark methods, i.e., offloading-only, local-computing-only and LR-Round, are sensitive to the network parameters, e.g., path loss exponent, placement, and network size, which may produce very poor performance in some practical setups. In contrast, regardless of the choice of initial condition, the proposed CD and ADMM methods can both achieve similar and superior computation rate performance under different network setups.

\begin{figure}
\centering
  \begin{center}
    \includegraphics[width=0.55\textwidth]{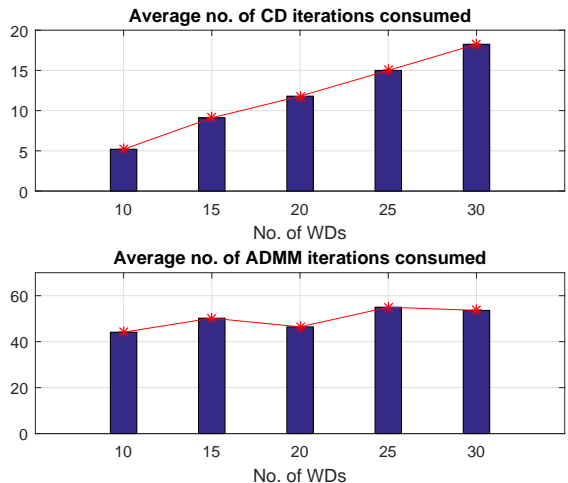}
  \end{center}
  \caption{Average number of iterations before convergence of the proposed CD (figure above) and ADMM (figure below) based methods when the number of WDs varies.}
  \label{106}
\end{figure}

\subsection{Computational Complexity Evaluation}
In Fig.~\ref{106}, we characterize the computational complexity of the proposed CD- and ADMM-based algorithms. Here, we use the same network setup as in Fig.~\ref{105} and examine the convergence rates of the two methods when $N$ increases. With the termination criterions in Algorithm $2$ and $3$, we plot the average number of iterations consumed by the CD and the ADMM-based methods before their convergence. Specifically, we observe that the iteration number of the CD method increases linearly with $N$, i.e., $O(N)$. Because each CD iteration runs Algorithm $1$ exactly $N$ times, the total number of executions of Algorithm $1$ scales as $O(N^2)$. Furthermore, because the computational complexity of Algorithm $1$ is $O(N)$, the overall time complexity of the CD method is $O(N^3)$. On the other hand, the ADMM-based method consumes almost constant number of iterations within the considered range of $N$, i.e., $O(1)$ complexity. Because each ADMM iteration is of $O(N)$ complexity, the overall computational complexity of the ADMM-based method is $O(N)$. The above results show that, although the computation time of the ADMM-based method could be longer than the CD method when $N$ is small, its computational complexity increases in a much slower pace than the CD method, i.e., $O(N)$ versus $O(N^3)$, thus is more manageable in a large-size IoT network (e.g., consisting of tens to several hundred of WDs) where the network size dominates the overall complexity.

\section{Conclusions and Future Work}
In this paper, we studied a weighted sum computation rate maximization problem in multi-user wireless powered edge computing networks with binary computation offloading policy. We formulated the problem as a joint optimization of individual computing mode selection and system transmission time allocation. In particular, we proposed two efficient solution algorithms to tackle the difficult combinatorial computing mode selection, where one coordinate descent method decouples the optimizations of mode selection and time allocation, and the other ADMM-based method optimizes them jointly. For a homogeneous special case, we observe an interesting threshold structure in the optimal computing mode solution based on wireless channel gain. Extensive simulation results showed that both the proposed CD-based and ADMM-based methods can achieve near-optimal computation rate performance under different network setups, and significantly outperform the other representative benchmark methods.

In practical implementation, the CD method requires only basic function evaluations, while the ADMM-based method needs to run more complex convex optimization algorithms. However, the ADMM-based method has a $O(N)$ computational complexity in network size $N$ compared to the $O(N^3)$ complexity of the CD method. Therefore, it is more preferable to use the CD method when network size is small or the MEC server is hardware-constrained, and to use ADMM-based method in large-scale networks where the network size dominates the overall complexity.

Finally, we conclude the paper with some interesting future working directions of wireless powered MEC. First, we assumed in this paper that the MEC server has unlimited computing capacity. In practice, massive offloading tasks may overwhelm the MEC server such that it needs to allocate its computing power among the offloading tasks received. As a result, the computation delay at the MEC server becomes non-negligible, thus should be jointly considered with task offloading time. Second, it is interesting to extend the problem to fading channels, such that a WD may choose to store the harvested energy in the battery in some time slots instead of performing immediate local computing or offloading. At last, it is also challenging to extend the considered network model to other practical setups, such as multi-antenna AP, relay channel, user cooperation, and interference channel, etc.

\begin{appendices}

\section{Proof of Lemma $2$}
\emph{Proof}: \ The partial derivative of $L$ with respect to $\tau_j$ is
\begin{equation}
\small
\begin{aligned}
\frac{\partial L}{\partial\tau_j} &=  w_j\varepsilon \ln \left(1+ \frac{\eta_2 h_j^2 a }{\tau_j}\right) - \frac{w_j\varepsilon  \cdot \eta_2 h_j^2 a \tau_j^{-1}}{1+ \eta_2 h_j^2 a \tau_j^{-1}} - \nu.
\end{aligned}
\end{equation}
By setting $\frac{\partial L}{\partial\tau_j}=0$ at the maximum point, we have
\begin{equation}
\small
\begin{aligned}
\ln\left(1+ \eta_2 h_j^2 a \tau_j^{-1}\right) &= (1+\frac{\nu}{w_j\varepsilon}) -  \frac{1}{1+\eta_2 h_j^2 a \tau_j^{-1}}.
\end{aligned}
\end{equation}
By taking a natural exponential operation at both sides, we have
\begin{equation}
\small
\label{34}
\left(1+ \eta_2 h_j^2 a \tau_j^{-1}\right) \exp\left(\frac{1}{1+ \eta_2 h_j^2 a \tau_j^{-1}}\right)= \exp\left(1+\frac{\nu}{w_j\varepsilon}\right).
\end{equation}

Consider two positive values $x$ and $z$ that satisfy $\frac{1}{x} \exp(x) = z$, it holds that
\begin{equation}
\label{35}
-x \exp(-x) = - \frac{1}{z}.
\end{equation}
Therefore, we have $x = -W(-\frac{1}{z})$, where $W(v)$ denotes the Lambert-W function, which is the inverse function of $f(u) = u \exp(u) =v$, i.e., $u = W(v)$. Comparing (\ref{34}) and (\ref{35}), it is straightforward to infer that $\frac{1}{1+ \eta_2 h_j^2 a \tau_j^{-1}} = -W\left(-\frac{1}{\exp(1+\frac{\nu}{w_j\varepsilon})}\right)$, which leads to the result in Lemma $1$ with some simple manipulation. $\hfill \blacksquare$

\section{Proof of Proposition $1$}
\emph{Proof}: Take the partial derivative of $L$ in (\ref{14}) with respect to $a$. The maximum of $L$ is achieved when
\begin{equation}
\label{15}
\begin{aligned}
\frac{\partial L}{\partial a} &=   \frac{1}{3} (a^*)^{-\frac{2}{3}}\sum_{i \in \mathcal{M}_0} w_i  \eta_1 \left(\frac{h_i}{k_i}\right)^{\frac{1}{3}} + \sum_{j\in \mathcal{M}_1} \frac{w_j\varepsilon \eta_2 h_j^2}{1+ \eta_2 h_j^2 a^* (\tau_i^*)^{-1}} - \nu = 0.
\end{aligned}
\end{equation}
From (\ref{10}), it holds that
\begin{equation}
\label{33}
\eta_2 h_j^2 a^* (\tau_i^*)^{-1} = \frac{1}{\varphi_j(\nu^*)}.
\end{equation}
By substituting (\ref{8}) and (\ref{33}) into (\ref{15}), we see that the optimal $\nu^*$ must satisfy
\begin{equation}
Q(\nu^*) \triangleq \frac{1}{3} \left(p_1(\nu^*)\right)^{-\frac{2}{3}}\sum_{i \in \mathcal{M}_0} w_i  \eta_1 \left(\frac{h_i}{k_i}\right)^{\frac{1}{3}} + \varepsilon \eta_2 \sum_{j\in \mathcal{M}_1} \frac{w_j h_j^2}{1+1/\varphi_j(\nu^*)} - \nu^* = 0.
\end{equation}
Now that (P2) is convex given $\mathcal{M}_0$, $Q(\nu^*)=0$ is a sufficient condition for optimality. We then show that such $\nu^*$ exists and is unique. Notice that $p_1(\nu)$ is an increasing function in $\nu$ and $\varphi_j(\nu)$ is a decreasing function in $\nu$. Therefore, all the three terms in $Q(\nu)$ decrease with $\nu$, thus $Q(\nu)$ is a monotonically decreasing function in $\nu$. Meanwhile, when $\nu \rightarrow 0$, it holds that $p_1(\nu)\rightarrow 0$ and $\varphi_j(\nu) \rightarrow \infty$. Thus, we have $Q(\nu)\rightarrow \infty$ when $\nu \rightarrow 0$. Besides, when $\nu \rightarrow \infty$, it holds that $p_1(\nu)\rightarrow 1$ and $\varphi_j(\nu) \rightarrow 0$, which leads to $Q(\nu)\rightarrow -\infty$ when $\nu \rightarrow \infty$. Together with the result that $Q(\nu)$ is a monotonically decreasing function, there must exist a unique $\nu^*>0$ that satisfies $Q(\nu^*)=0$ at the optimum. This completes the proof of Proposition $1$. $\hfill \blacksquare$

\end{appendices}

\end{document}